\documentclass[aps,prd,superscriptaddress,showpacs]{revtex4}
\usepackage{graphicx}
\usepackage{epstopdf}
\usepackage{amsmath}
\usepackage{amsfonts}
\usepackage{amssymb}
\usepackage{latexsym}

\begin{document}
\title{On finite interquark potential in $D=3$ driven by a minimal length}
\author{Patricio Gaete}\email{patricio.gaete@usm.cl}
 \affiliation{Departmento de F\'{i}sica and Centro Cient\'{i}fico-Tecnol\'ogico de Valpara\'{i}so, Universidad T\'{e}cnica Federico Santa Mar\'{i}a, Valpara\'{i}so, Chile}

\begin{abstract}
We address the effect of a quantum gravity induced minimal length on a physical observable for three- dimensional Yang-Mills. Our calculation is done within stationary perturbation theory. Interestingly enough, we find an ultraviolet finite interaction energy, which contains a regularized logarithmic function and a linear confining potential. This result highlights the role played by the new quantum of length in our discussion.
\end{abstract}
\maketitle

\section{Introduction}

It is known that one of the main unsolved problems in high energy physics is a quantitative description (from first principles) of confinement in quantum chromodynamics (QCD). Albeit phenomenological models still represent a key tool for understanding confinement physics. In this context we recall the phenomenon of condensation, where in the scenario of dual superconductivity, it is conjectured that the QCD vacuum behaves as a dual-type II superconductor. More explicitly, due to the condensation of magnetic monopoles, the chromo-electric field acting between $q\bar q$ pair is squeezed into strings, and the non vanishing string tension represents the proportionality constant in the linear potential. Incidentally, lattice calculations have confirmed this picture by showing the formation of tubes of gluonic fields connecting colored charges \cite{Capstick}.

It is also known that considerable attention has been paid to the investigation of extensions of the Standard Model (SM), such as Lorentz invariance violation and fundamental length
{\cite{AmelinoCamelia:2002wr,Jacobson:2002hd,Konopka:2002tt,Hossenfelder:2006cw,Nicolini:2008aj,Euro}}, because the SM does not include a quantum theory of gravitation. In this respect we recall that, in the last few years, the emphasis of quantum gravity has been on effective models incorporating a minimal length scale. In fact, there are several approaches on how to incorporate a minimal length scale in a quantum field theory, leading to a model of quantum space-time
\cite{Magueijo:2001cr,Hossenfelder:2003jz,Nicolini:2010bj,Sprenger:2012uc}.
Of these, non-commutative quantum field theories have motivated a great interest
\cite{Witten:1985cc,Seiberg:1999vs,Douglas:2001ba,Szabo:2001kg,Gomis:2000sp,Bichl:2001nf}. Notice that this non-commutative geometry is an intrinsic property of space-time. In addition, we also recall that most of the known results in the non-commutative approach have been achieved using a
Moyal star-product. Nevertheless, in recent times, a new formulation
of  non-commutative quantum field theory in the presence of a
minimal length has been proposed in
\cite{Smailagic:2003rp,Smailagic:2003yb,Smailagic:2004yy}.
Afterwards, this approach was further developed by the introduction of a new
multiplication rule, which is known as Voros star-product. 
Notwithstanding, physics turns out be independent from the choice of the
type of product \cite{Hammou:2001cc}. Consequently, with the
introduction of non-commutativity by means of a minimal length, the
theory becomes ultraviolet finite and the cutoff is provided by the non-commutative parameter $\theta$.

In this perspective the present work is an extension of our previous study \cite{Gaete13}. Thus, the basic ideas underlying the analysis of this paper are derived from our earlier paper \cite{Gaete13}. Specifically, in this work we will focus attention on the impact of a minimal length on a physical observable for pure Yang-Mills theory in $(2+1)$D. It is worthy noting here that Yang-Mills theories in $(2+1)$D are very relevant for a reliable comparison between results coming from continuum and lattice calculations \cite{Feuchter}. Also, $(2+1)$D theories have been raising a great deal of interest in connection with branes activity, for example: issues like self-duality \cite{Singh}, and new possibilities for supersymmetry breaking as induced by $3$-branes \cite{Bergman}. Yet, $(2+1)$D theories may be adopted to describe the high-temperature limit of models in $(3+1)$D \cite{Das}. In fact, such theories are of interest to probe low-dimensional condensed matter systems, such as spin or pairing fluctuations by means of effective gauge theories, for which $(2+1)$D theories are a very good approximation \cite{Klevshenko}.
Thus, in order to accomplish the purpose of studying the impact of a minimal length for Yang-Mills theory in $(2+1)$D, we shall work out the static
potential for the case under consideration. As we shall see, the presence of a
minimal length leads to an ultraviolet finite static potential, which contains a regularized logarithmic function and a linear confining potential. Accordingly, our study offers a straightforward calculation in which some features of three-dimensional non-Abelian gauge theories become more transparent.

\section{Interaction energy}

We shall now discuss the interaction energy for Yang-Mills theory
in the presence of a minimal length. We start then with the three-dimensional space-time Lagrangian:
\begin{eqnarray}
{\cal L} =  - \frac{1}{4}Tr\left( {F_{\mu \nu } F^{\mu \nu } } \right) 
=  - \frac{1}{4}F_{\mu \nu }^a F^{a\mu \nu }.
\label{3DYM05}
\end{eqnarray}
Here $ A_\mu  \left( x \right) = A_\mu ^a \left( x \right)T^a $ and
$ F_{\mu \nu }^a  = \partial _\mu  A_\nu
^a - \partial _\nu  A_\mu ^a  + gf^{abc} A_\mu ^b A_\nu ^c$, with
$f^{abc}$ the structure constants of the gauge group. 

As we have indicated in \cite{Gaete13}, our analysis is based in perturbation theory along the lines of Refs. \cite{Drell, Gribov, Lavelle}. To do this, we will work out the vacuum expectation value of the energy operator $H$ $(\left\langle 0 \right|H\left| 0 \right\rangle)$ at lowest order in $g$, in the Coulomb gauge. The canonical Hamiltonian can be worked as usual and is given by 
\begin{equation}
H = \frac{1}{2}\int {{d^2}x} \left\{ {{{\left( {{\bf E}_T^a} \right)}^2} +
{{\left( {{{\bf B}^a}} \right)}^2} - {\phi ^a}{\nabla ^2}{\phi ^a}} \right\}, 
\label{3DYM10}
\end{equation}
where the color-electric field ${{\bf E}^a }$ has been separated into transverse and longitudinal parts: ${{\bf E}^a} = {\bf E}_T^a - \nabla {\phi ^a}$. 

Next, by making use of Gauss's law
\begin{equation}
{\nabla ^2}{\phi ^a} = g\left( {{\rho ^a} - {f^{abc}}{{\bf A}^b} \cdot {{\bf E}^c}} \right), \label{3DYM15}
\end{equation}
we get
\begin{equation}
{\nabla ^2}{\phi ^a} = \left( {g{\delta ^{ap}} + {g^2}{f^{abp}}{{\bf A}^b} \cdot {\bf\nabla} \frac{1}{{{\nabla ^2}}} + {g^3}{f^{abc}}{{\bf A}^b} \cdot \nabla \frac{1}{{{\nabla ^2}}}{f^{chp}}{{\bf A}^h} \cdot \nabla \frac{1}{{{\nabla ^2}}}} \right) 
\left( {{\rho ^p} - {f^{pde}}{{\bf A}^d} \cdot {\bf E}_T^e} \right). \label{3DYM20}
\end{equation}

Following our earlier procedure, the corresponding formulation of this theory in the presence of a minimal length is by means of a smeared source \cite{Gaete13,Gaete:2011ka,GaeteSpaHel}. Thus, we will take the sources as $\rho ^a  \equiv \rho _1^a  + \rho _2^a  = \rho _{\bar q}^a  + \rho _q^a $, where $\rho _{\bar q}^a({\bf x})  = t_{\bar q}^a e^{{\raise0.5ex\hbox{$\scriptstyle \theta $}\kern-0.1em/\kern-0.15em
\lower0.25ex\hbox{$\scriptstyle 2$}}\nabla ^2 } \delta ^{\left( 3 \right)} \left( {{\bf x} - {\bf y} \prime } \right)$ and $
\rho _q^a({\bf x}) = t_q^a e^{{\raise0.5ex\hbox{$\scriptstyle \theta $}
\kern-0.1em/\kern-0.15em
\lower0.25ex\hbox{$\scriptstyle 2$}}\nabla ^2 } \delta ^{\left( 3 \right)} \left( {{\bf x} - {\bf y}} \right)$, where $t_{\bar q}^a$ and $t_q^a$ are the color charges of a heavy antiquark ${\bar q}_i$ and a quark $q_i$ in a normalized color singlet state $  
\left| \Psi  \right\rangle  = N^{ - {\raise0.5ex\hbox{$\scriptstyle 1$}
\kern-0.1em/\kern-0.15em
\lower0.25ex\hbox{$\scriptstyle 2$}}} \left| {q_i } \right\rangle \left| {\bar q_i } \right\rangle$. Hence $
t_q^a t_{\bar q}^a  = {\textstyle{1 \over N}}tr\left( {T^a T^a } \right) =  - C_F$, where the anti-Hermitian generators $T^a$ are in the fundamental representation of SU(N).

By proceeding in the same way as in \cite{Gaete13}, we obtain the expectation value of the energy operator $H$ to order $g^2$  and $g^4$:
\begin{equation}
V=V_1 +V_2, \label{3DYM25}
\end{equation}
where
\begin{equation}
V_1 =  - {g^2}\int {{d^2}x} \left\langle 0 \right|{\rho _1^a}\frac{1}{{{\nabla ^2}}}{\rho _2^a}\left| 0 \right\rangle, \label{3DYM30}
\end{equation}
and
\begin{equation}
V_2 =- 3{g^4}f^{abc}f^{chq} 
\int {{d^2}x} \left\langle 0 \right|{\rho _1^a}\frac{1}{{{\nabla ^2}}}{{\bf A}^b} \cdot \nabla \frac{1}{{{\nabla ^2}}}{{\bf A}^h} \cdot \nabla \frac{1}{{{\nabla ^2}}}{\rho _2^q}\left| 0 \right\rangle. \label{3DYM35}
\end{equation}

The $V_1$ term is exactly the one obtained in \cite{GaeteSpaHel}. Consequently, Eq. ($\ref{3DYM30}$) takes the form
\begin{eqnarray}
V_1  &=&  - g^2 C_F \int {\frac{{d^2 k}}{{\left( {2\pi } \right)^2 }}} \frac{{e^{ - \theta {\bf k}^2 } }}{{{\bf k}^2 }}e^{ - i{\bf k} \cdot {\bf r}} \nonumber\\ 
&=&  
\frac{{g^2 C_F }}{{2\pi }}\left\{ {\ln \left( {\mu r} \right) - e^{ - {\textstyle{{r^2 } \over {4\theta }}}} \ln \left( {\frac{r}{{2\sqrt \theta  }}} \right)} \right\}, \label{3DYM40}
\end{eqnarray}
with $|{\bf r}|\equiv |{\bf y}-{\bf y}^\prime|= r$ and $\mu$ is an infrared regulator. Again, as in our previous analysis \cite{GaeteSpaHel}, unexpected features are found. Interestingly, it is observed that unlike the Coulomb potential which is singular at the origin, $V_1$ is finite there:  
$V_1  = \frac{{g^2 C_F }}{{2\pi }}\ln \left( {2\mu \sqrt \theta  } \right)$.

We now turn our attention to the $V_2$ term, which is given by
\begin{equation}
V_2  = 3g^4 C_A C_F \int {\frac{{d^2 k}}{{\left( {2\pi } \right)^2 }}} \frac{{e^{ - \theta {\bf k}^2 } }}{{{\bf k}^2 }}e^{ - i{\bf k} \cdot {\bf r}} {\cal I}({\bf k}), \label{3DYM45}
\end{equation}
where
\begin{equation}
{\cal I}({\bf k})=\int {\frac{{d^2 p}}{{\left( {2\pi } \right)^2 }}} \frac{1}{{2|{\bf p}|\left( {{\bf p} - {\bf k}} \right)^2 }}\left( {1 - \frac{{\left( {{\bf p} \cdot {\bf k}} \right)^2 }}{{{\bf p}^2 {\bf k}^2 }}} \right). \label{3DYM50}
\end{equation}
In passing we recall that to obtain Eq.(\ref{3DYM45}) we have expressed the $A^{ai}$-fields in terms of a normal mode expansion: $    
A^{ai} ({\bf x},t) = \int {\frac{{d^3 p}}{{\sqrt {\left( {2\pi } \right)^3 2w_{\bf p} } }}} \sum\limits_\lambda  {\varepsilon ^i } \left( {{\bf p},\lambda } \right)\left[ {a^a \left( {{\bf p},\lambda } \right)e^{ - ipx}  + a^{\dag a} \left( {{\bf p},\lambda } \right)e^{ipx} } \right]$, along with $    
\left[ {a^a \left( {{\bf p},\lambda } \right),a^{\dag b} \left( {{\bf l},\sigma } \right)} \right] = \delta ^{ab} \delta ^{\lambda \sigma } \delta ^{\left( 3 \right)} \left( {{\bf p} - {\bf l}} \right)$ and $  
\sum\limits_\lambda  {\varepsilon ^i \left( {{\bf k},\lambda } \right)} \varepsilon ^j \left( {{\bf k},\lambda } \right) = \delta ^{ij}  - \frac{{k^i k^j }}{{{\bf k}^2 }}$. Here we would mention that the correction term of order $g^4$ represents an anti-screening effect. Incidentally, it is of interest to notice that precisely this term is in the origin of asymptotic freedom in the $(3+1)$D case, which is due to the instantaneous Coulomb interaction of the quarks.

When the integral (\ref{3DYM50}) is performed, one gets 
\begin{equation}
{\cal I}\left( k \right) = \frac{1}{{8\pi ^2 }}\frac{1}{{|{\bf k}|}}\left( { - 1.5706 + \frac{\pi }{2}} \right). \label{3DYM55}
\end{equation}
Expression (\ref{3DYM45}) then becomes 
\begin{equation}
V_2  =  - \frac{3}{{8\pi ^2 }}\left( {1.5706 - \frac{\pi }{2}} \right)g^4 C_A C_F \int {\frac{{d^2 k}}{{\left( {2\pi } \right)^2 }}} \frac{{e^{ - \theta {\bf k}^2 } }}{{|{\bf k}|^3 }}e^{ - i{\bf k} \cdot {\bf r}}. \label{3DYM60}
\end{equation}

We now proceed to calculate the integral (\ref{3DYM60}). Following our earlier procedure \cite{GaeteSpaHel}, equation (\ref{3DYM60}) is further rewritten as 
\begin{eqnarray}
I \equiv \mathop {\lim }\limits_{\varepsilon  \to 0} \tilde I &=& \mathop {\lim }\limits_{\varepsilon  \to 0} \left( {\mu ^2 } \right)^{ - {\raise0.5ex\hbox{$\scriptstyle \varepsilon $}
\kern-0.1em/\kern-0.15em
\lower0.25ex\hbox{$\scriptstyle 2$}}} \int {\frac{{d^{2 + \varepsilon } k}}{{\left( {2\pi } \right)^2 }}} \frac{{e^{ - \theta {\bf k}^2 } }}{{|k|^3 }}e^{i{\bf k} \cdot {\bf r}} , \nonumber\\
&=& \mathop {\lim }\limits_{\varepsilon  \to 0} \frac{{\left( {\mu ^2 } \right)^{ - {\raise0.5ex\hbox{$\scriptstyle \varepsilon $}
\kern-0.1em/\kern-0.15em
\lower0.25ex\hbox{$\scriptstyle 2$}}} }}{{\Gamma \left( {{\raise0.5ex\hbox{$\scriptstyle 3$}
\kern-0.1em/\kern-0.15em
\lower0.25ex\hbox{$\scriptstyle 2$}}} \right)}}\int_0^\infty  {dx}\,  x^{{\raise0.5ex\hbox{$\scriptstyle 1$}
\kern-0.1em/\kern-0.15em
\lower0.25ex\hbox{$\scriptstyle 2$}}} 
\int {\frac{{d^{2 + \varepsilon } k}}{{\left( {2\pi } \right)^2 }}e^{ - \left( {\theta  + x} \right){\bf k}^2 } } e^{i{\bf k} \cdot {\bf r}} .  \label{3DYM65}
\end{eqnarray}
Then, the $I$ term takes the form 
\begin{equation}
 I = \mathop {\lim }\limits_{\varepsilon  \to 0} \frac{{\left( {\mu ^2 } \right)^{ - {\raise0.5ex\hbox{$\scriptstyle \varepsilon $}
\kern-0.1em/\kern-0.15em
\lower0.25ex\hbox{$\scriptstyle 2$}}} \left( {r^2 } \right)^{^{ - {\raise0.5ex\hbox{$\scriptstyle \varepsilon $}
\kern-0.1em/\kern-0.15em
\lower0.25ex\hbox{$\scriptstyle 2$}}} } }}{{\Gamma \left( {{\raise0.5ex\hbox{$\scriptstyle 3$}
\kern-0.1em/\kern-0.15em
\lower0.25ex\hbox{$\scriptstyle 2$}}} \right)\left( \pi  \right)^{1 + {\raise0.5ex\hbox{$\scriptstyle \varepsilon $}
\kern-0.1em/\kern-0.15em
\lower0.25ex\hbox{$\scriptstyle 2$}}} }} 
\frac{1}{4}\int_0^{{\raise0.5ex\hbox{$\scriptstyle {r^2 }$}
\kern-0.1em/\kern-0.15em
\lower0.25ex\hbox{$\scriptstyle {4\theta }$}}} {d\tau } \tau ^{{\textstyle{\varepsilon  \over 2}} - 1} e^{ - \tau } \left( {\frac{{r^2 }}{{4\tau }} - \theta } \right)^{{\raise0.7ex\hbox{$1$} \!\mathord{\left/
 {\vphantom {1 2}}\right.\kern-\nulldelimiterspace}
\!\lower0.7ex\hbox{$2$}}} . \label{3DYM70}
\end{equation}
Hence, at leading order in $\theta$, equation (\ref{3DYM70}) reduces to
\begin{equation}
 I =  - \frac{1}{{2\left( \pi  \right)^{{\raise0.5ex\hbox{$\scriptstyle 3$}
\kern-0.1em/\kern-0.15em
\lower0.25ex\hbox{$\scriptstyle 2$}}} }} 
\left\{ {r\,\gamma \left( {{\textstyle{1 \over 2}},{\textstyle{{r^2 } \over {4\theta }}}} \right) + 2\sqrt \theta\,e^{ - {\textstyle{{r^2 } \over {4\theta }}}}  + \frac{\theta }{r}\,\gamma \left( {{\textstyle{1 \over 2}},{\textstyle{{r^2 } \over {4\theta }}}} \right)} \right\}, \label{3DYM75}
\end{equation}
where    
$\gamma \left( {{\textstyle{1 \over 2}},{\textstyle{{r^2 } \over {4\theta }}}} \right)
$ is the lower incomplete Gamma function defined by the following integral representation
\begin{equation}
\gamma \left( {{\textstyle{a \over b}}}, x \right) \equiv \int_0^x {\frac{{du}}{u}} u^{{\textstyle{a \over b}}} e^{ - u}. \label{3DYM80}
\end{equation}

Accordingly, the $V_2$ term reads 
\begin{equation}
V_2  = \frac{3}{{16\pi ^{{\raise0.5ex\hbox{$\scriptstyle 7$}
\kern-0.1em/\kern-0.15em
\lower0.25ex\hbox{$\scriptstyle 2$}}} }}\left( {1.5706 - \frac{\pi }{2}} \right)g^4 C_A C_F 
\left\{ {r\,\gamma \left( {{\textstyle{1 \over 2}},{\textstyle{{r^2 } \over {4\theta }}}} \right) + 2\sqrt \theta\,e^{ - {\textstyle{{r^2 } \over {4\theta }}}}  + \frac{\theta }{r}\,\gamma \left( {{\textstyle{1 \over 2}},{\textstyle{{r^2 } \over {4\theta }}}} \right)} \right\}. \nonumber\\
\label{3DYM85}
\end{equation}

Now we focus on the $(g^4)$ screening contribution to the potential, which is due to the exchange of transverse gluons. From our above perturbation theory, we find that $V_2^ *$ is given by
\begin{equation}
V_2^ * = 2g^4 f_{abc} f_{def} \sum\limits_{n = 2 \ gluon} {\frac{1}{{E_n }}} \int {d^3 x} \int {d^3 w} \left\langle 0 \right|\rho _2^a \frac{1}{{\nabla ^2 }}{\bf A}^b \cdot {\bf E}_T^c \left| n \right\rangle _{\bf x} 
\left\langle n \right|\rho _1^d \frac{1}{{\nabla ^2 }}{\bf A}^e \cdot {\bf E}_T^f \left| 0 \right\rangle _{\bf w}. 
\label{3DYM90} 
\end{equation}
In passing we recall that, $\left\langle 0 \right|\rho _2^a \frac{1}{{\nabla ^2 }}{\bf A}^b \cdot {\bf E}_T^c \left| n \right\rangle$, is the matrix element in the basis of states in which the non perturbated Hamiltonian is diagonal. Next, it should be noted that the intermediate state $\left| n \right\rangle$ must contains a pair of transverse gluons, since the terms $ {\bf A}^b \cdot {\bf E}_T^c $ must create and destroy dynamical gluon pairs. We can, therefore, write two gluon states as 
\begin{equation}
\sum\limits_{n = 2 \ gluon} {\left| n \right\rangle } \left\langle n \right| = \frac{1}{2}\sum\limits_{kl}\sum\limits_{\lambda \sigma } {\int {d^3 k} } \int {d^3 } l\  
a^{\dag e} \left( {{\bf  k},\lambda } \right)a^{\dag f} \left( {{\bf    l},\sigma } \right)\left| 0 \right\rangle 
\left\langle 0 \right|a^f \left( {{\bf  l},\sigma } \right)a^e \left( {{\bf  k},\lambda } \right). 
\label{3DYM90b}
\end{equation}
By substituting Eq. (\ref{3DYM90b}) into Eq. (\ref{3DYM90}) and following our earlier procedure, the $ V_2^ *$ term assumes the form
\begin{equation}
V_2^ *   =  - C_A C_F g^4 \int {\frac{{d^2 k}}{{\left( {2\pi } \right)^2 }}} \frac{{e^{ - \theta {\bf k}^2 } }}{{{\bf k}^4 }}e^{i{\bf k} \cdot {\bf r}} {\cal I}\left( {\bf k} \right), \label{3DYM95}
\end{equation}
where
\begin{equation}
{\cal I}\left( {\bf k} \right) = \int {\frac{{d^2 l}}{{\left( {2\pi } \right)^2 }}} \frac{{\left( {|{\bf l}| - |{\bf l} - {\bf k}|} \right)^2 }}{{4|{\bf l}||{\bf l} - {\bf k}|\left( {|{\bf l}| + |{\bf l} - {\bf k}|} \right)}} 
\left( {1 - \frac{{{\bf k}^2 }}{{\left( {{\bf l} - {\bf k}} \right)^2 }} + \frac{{\left( {{\bf k} \cdot {\bf l}} \right)^2 }}{{\left( {{\bf l} - {\bf k}} \right)^2 {\bf l}^2 }}} \right)
. \label{3DYM100}
\end{equation}

Integrating now over ${\bf l}$, one then obtains 
${\cal I}\left( {\bf k} \right) = |{\bf k}|\left( {\frac{1}{{2\pi }}\frac{{15}}{{16}} - \frac{1}{8} + \frac{3}{{32\pi ^2 }}} \right)$. As a consequence, the $V_2^ *$ term becomes
\begin{equation}
V_2^ *  =  - \left( {\frac{1}{{2\pi }}\frac{{15}}{{16}} - \frac{1}{8} + \frac{3}{{32\pi ^2 }}} \right)g^4 C_A C_F 
\int {\frac{{d^2 k}}{{\left( {2\pi } \right)^2 }}} \frac{{e^{ - \theta {\bf k}^2 } }}{{|{\bf k}|^3 }}e^{ - i{\bf k} \cdot {\bf r}} . \label{3DYM105}
\end{equation}
It is straightforward to see that this integral is exactly the one obtained in expression (\ref{3DYM60}).

By putting together Eqs. (\ref{3DYM40}), (\ref{3DYM60}) and (\ref{3DYM105}), we evaluate the interquark potential in position space. We thus finally obtain 
\begin{eqnarray}
V(r) &=& \frac{{g^2 C_F }}{{2\pi }}\left\{ {\ln \left( {\mu r} \right) - e^{ - {\raise0.7ex\hbox{${r^2 }$} \!\mathord{\left/
 {\vphantom {{r^2 } {4\theta }}}\right.\kern-\nulldelimiterspace}
\!\lower0.7ex\hbox{${4\theta }$}}} \ln \left( {\frac{r}{{2\sqrt \theta  }}} \right)} \right\} \nonumber\\
&+& \frac{{g^4 C_A C_F }}{{2\left( \pi  \right)^{{\raise0.5ex\hbox{$\scriptstyle 3$}
\kern-0.1em/\kern-0.15em
\lower0.25ex\hbox{$\scriptstyle 2$}}} }}0.165
  \left\{ {r\,\gamma \left( {{\textstyle{1 \over 2}},{\textstyle{{r^2 } \over {4\theta }}}} \right) + 2\sqrt \theta\,  e^{ - {\textstyle{{r^2 } \over {4\theta }}}}  + \frac{\theta }{r}\,\gamma \left( {{\textstyle{1 \over 2}},{\textstyle{{r^2 } \over {4\theta }}}} \right)} \right\}, \nonumber\\
\label{3DYM110}
\end{eqnarray}
which is ultraviolet finite (Fig. 1).
An immediate consequence of this is that for $\theta=0$ one obtains the known interquark potential at order $g^4$ \cite{Lavelle}. Note that in figure $1$, for illustrative purposes, we have defined $
\frac{{g^2 C_F }}{{2\pi }} = 1$, $  
\frac{{g^4 C_A C_F }}{{2\left( \pi  \right)^{{\raise0.5ex\hbox{$\scriptstyle 3$}
\kern-0.1em/\kern-0.15em
\lower0.25ex\hbox{$\scriptstyle 2$}}} }} = 1$, $\mu  = 1$ and $\sqrt \theta   = 1.4$.

\begin{figure}[h]
\begin{center}
\includegraphics[scale=1.10]{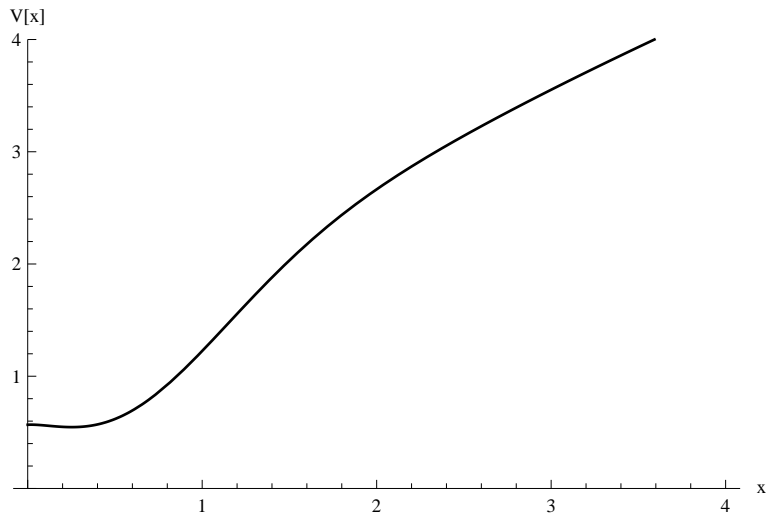}
\end{center}
\caption{\small The potential $V$, as a function of $x
= {\textstyle{r \over {2\sqrt \theta  }}}$.
\label{fig1}}
\end{figure}

\section{Conclusion}

To conclude, let us put our work in its proper perspective. As already anticipated, this work is a sequel to \cite{Gaete13}, where we have considered a three-dimensional extension of the recently $(3+1)$D calculation in the presence of a minimal length. To do this, we have exploited a correct identification of field degrees with observable quantities. Once the identification has been made, the computation of the potential is achieved by means of Gauss' law.
Interestingly enough, it was found that the static potential profile is ultraviolet finite, which contains a regularized logarithmic function and a linear potential
leading to confinement of static sources. Finally, we note that our results agree for the $\theta=0$ case with the calculation shown in \cite{Lavelle}. Also very recently, in the context of the Georgi-Glashow model it has been shown that there is confinement at distances much larger than the screening length \cite{Anber}. Since our calculation has shown that there is confinement in three-dimensional Yang-Mills, it seems a challenging work to extend the above analysis to the Georgi-Glashow model. We expect to report on progress along this lines soon.

\begin{acknowledgments}
I would like to thank Jos\'e A. Helay\"el-Neto for useful discussions. This work was partially supported by Fondecyt (Chile) Grant 1130426. He author also wishes to thank the Field Theory Group of the CBPF for hospitality and PCI/MCT for support.
\end{acknowledgments}

\end{document}